\documentclass[fleqn,10pt]{wlscirep}
\usepackage[utf8]{inputenc}
\usepackage[T1]{fontenc}

\RequirePackage{fix-cm}
\makeatletter\if@twocolumn\PassOptionsToPackage{switch}{lineno}\else\fi\makeatother

  \makeatletter
  \def\fig@textbf{\textbf}
   \AtBeginDocument{\renewcommand\floatc@plain[2]{\setbox\@tempboxa\hbox{\textbf{\footnotesize#1.}\footnotesize\hskip.5em#2}%
    \ifdim\wd\@tempboxa>\hsize {\fig@textbf{\footnotesize#1.}}\footnotesize\hskip.5em#2\par
        \else\hbox to\hsize{\hfil\box\@tempboxa\hfil}\fi}}
    \makeatother
  
\usepackage{amsfonts,amssymb,amsbsy,tabulary,amsmath}
\usepackage{hyperref}
\usepackage{graphicx}
\usepackage{url,multirow,morefloats,floatflt,cancel,tfrupee}
\makeatletter

\AtBeginDocument{\@ifpackageloaded{textcomp}{}{\usepackage{textcomp}}}
\makeatother
\usepackage{colortbl}
\usepackage{xcolor}
\usepackage{pifont}
\usepackage[nointegrals]{wasysym}
\makeatletter

\def\mcWidth#1{\csname TY@F#1\endcsname+\tabcolsep}

\def\cAlignHack{\rightskip\@flushglue\leftskip\@flushglue\parindent\z@\parfillskip\z@skip}
\def\rAlignHack{\rightskip\z@skip\leftskip\@flushglue \parindent\z@\parfillskip\z@skip}

\@ifundefined{etal}{}{}

\usepackage{ifxetex}
\ifxetex\else\if@twocolumn\@ifpackageloaded{stfloats}{}{\usepackage{dblfloatfix}}\fi\fi

\AtBeginDocument{
\expandafter\ifx\csname eqalign\endcsname\relax
\def\eqalign#1{\null\vcenter{\def\\{\cr}\openup\jot\m@th
  \ialign{\strut$\displaystyle{##}$\hfil&$\displaystyle{{}##}$\hfil
      \crcr#1\crcr}}\,}
\fi
}

\AtBeginDocument{%
  \@ifpackageloaded{endfloat}%
   {\renewcommand\efloat@iwrite[1]{\immediate\expandafter\protected@write\csname efloat@post#1\endcsname{}}}{\newif\ifefloat@tables}%
}%

\def\BreakURLText#1{\@tfor\brk@tempa:=#1\do{\brk@tempa\hskip0pt}}
\let\lt=<
\let\gt=>
\def\processVert{\ifmmode|\else\textbar\fi}

\@ifundefined{subparagraph}{
\def\subparagraph{\@startsection{paragraph}{5}{2\parindent}{0ex plus 0.1ex minus 0.1ex}%
{0ex}{\normalfont\small\itshape}}%
}{}

\newcommand\role[1]{\unskip}
\newcommand\aucollab[1]{\unskip}
  
\@ifundefined{tsGraphicsScaleX}{\gdef\tsGraphicsScaleX{1}}{}
\@ifundefined{tsGraphicsScaleY}{\gdef\tsGraphicsScaleY{.9}}{}
\def\checkGraphicsWidth{\ifdim\Gin@nat@width>\linewidth
	\tsGraphicsScaleX\linewidth\else\Gin@nat@width\fi}

\def\checkGraphicsHeight{\ifdim\Gin@nat@height>.9\textheight
	\tsGraphicsScaleY\textheight\else\Gin@nat@height\fi}

\def\fixFloatSize#1{}
\let\ts@includegraphics\includegraphics

\def\inlinegraphic[#1]#2{{\edef\@tempa{#1}\edef\baseline@shift{\ifx\@tempa\@empty0\else#1\fi}\edef\tempZ{\the\numexpr(\numexpr(\baseline@shift*\f@size/100))}\protect\raisebox{\tempZ pt}{\ts@includegraphics{#2}}}}

\AtBeginDocument{\def\includegraphics{\@ifnextchar[{\ts@includegraphics}{\ts@includegraphics[width=\checkGraphicsWidth,height=\checkGraphicsHeight,keepaspectratio]}}}

\DeclareMathAlphabet{\mathpzc}{OT1}{pzc}{m}{it}

\def\URL#1#2{\@ifundefined{href}{#2}{\href{#1}{#2}}}

\def\UrlOrds{\do\*\do\-\do\~\do\'\do\"\do\-}%
\g@addto@macro{\UrlBreaks}{\UrlOrds}

\edef\fntEncoding{\f@encoding}

\makeatother

\newif\ifmultipleabstract\multipleabstractfalse%
\usepackage[T1]{fontenc}

\makeatletter	

\AtBeginDocument{%
  \@ifpackageloaded{endfloat}%
   {
\renewcommand*\efloat@process[2]{%
  \ef@ifct{#1}{%
    \expandafter\immediate\expandafter\closeout\csname efloat@post#1\endcsname
    \ef@setct{#1}{0}%
    \clearpage                                                         
        
    \efloat@ifflag{#2list}{
      {\normalsize\efloat@listof{#2}}
    }{}%

    \efloat@ifflag{#2head}{%
      \section*{\@nameuse{#2section}}
      \suppressfloats[t]
    }{}

    \markboth                                                          
      {\expandafter\uppercase\expandafter{\csname #2section\endcsname}}
      {\expandafter\uppercase\expandafter{\csname #2section\endcsname}}

    \def\efloat@type{#2}%
    \processdelayedfloat@hook
    \@nameuse{process#2s@hook}%
    \clearpage
    \@input{\jobname.#1}%
  }{}}
   
   }{}%
}%

  \makeatother

\usepackage[section]{placeins}

\usepackage{float}

\usepackage{listings,xcolor}

\lstdefinestyle{listing_style}{frame=single,basicstyle=\fontfamily{pcr}\selectfont,numberstyle=\tiny,xleftmargin=1pc,linewidth=.98\linewidth,backgroundcolor=\color{black!0},breaklines=true,keywordstyle=\color{blue},commentstyle=\color{darkgray},numbers=left,tabsize=3,captionpos=b,escapeinside={[@}{@]}}

\nocite{*}

\title{Analyses of the viability of automating the quantum circuit construction of Grover's Oracle for executing wildcard searches on NISQ processors}

\author[1]{Willie Huang}
\affil[1]{Amazon Web Services, Seattle, United States, willhuan@amazon.com}

\begin{abstract}
Using Grover's algorithm, this work investigates a technique for encoding search phrases used in wildcard searches.  The technique involves creating a phase Oracle that loads data into a quantum circuit together with the search terms that have been encoded.  The method entails constructing a phase Oracle programmatically using encoded input information and encoded search terms.  By combining Grover's diffusion operator with the phase Oracle, Hadamard gates, and zero-initialized three-qubit states, a complete quantum circuit is created. Trapped ion and superconducting qubit quantum computers, which were used in the research, were operated numerous times. In order to confirm that the proposed strategy is a workable one for wildcard search, the outputs from both systems were compared with the expected values. The suggested strategy will be useful for a range of wildcard search issues and could speed up the attainment of quantum advantage.\mbox{~}
\end{abstract}

\begin{document}

\flushbottom
\maketitle
%
%
\thispagestyle{empty}

\section*{Introduction}

\subsection*{Problem scope}Unstructured search performance optimization has become a popular area of study in recent years \unskip~\cite{1544700:26175382}.  This research will concentrate on examining the viability of programmatic data and search term loading into quantum circuits, followed by wildcard search. Moreover, the purpose of this study is to test the viability of the approach for simultaneously running wildcard search terms on NISQ (Noisy Intermediate-Scale Quantum \unskip~\cite{1544700:26156282} processors.

Prefix search, suffix search, and substring search would be the three subcategories of the wildcard search problem space. The substring search is a more general type of search that may also include prefix and suffix searches in addition to looking for alphanumeric characters within a word.  There are contemporary algorithms and system designs to enhance the performance of wildcard search in classical computers, although there may be a trade-off between runtime and space complexity. Additionally, The current process included searching for terms one at a time, which caused runtime to scale linearly with the quantity of terms in the database.

Furthermore, since the area of interest is wildcard search, doing an unstructured search would result in a runtime complexity of $\mathcal{O}(NMO)$, where $N$ is the length of the data set's items and $M$ is the length of the items that match the wildcard pattern. The worst-case scenario for $M$ is equal to $N$, whereas the best-case scenario for $M$ is $1$. The length of the wildcard search phrases is $O$.  A situation where the runtime increases in cubic powers as the data set and search terms both expand linearly would not be unheard of in wildcard searches. Grover's algorithm may offer a chance to exponentially enhance the runtime\unskip~\cite{1544700:26156610} of the aforementioned problem space.

\subsection*{ Background}Grover's algorithm was developed for unstructured search in large data sets in mind \unskip~\cite{1544700:26156567} . In comparison to a classical computer, which would need an average runtime of $N/2$ \unskip~\cite{1544700:26156610}, where $N$ is the size of the data set, the approach gives quantum computing an exponential speedup.  Among the more general amplitude amplification methods\unskip~\cite{1544700:26156873}.\unskip~\cite{1544700:26175287}, the Grover's algorithm is the most effective approach\unskip~\cite{1544700:26156741}\unskip~\cite{1544700:26175282}\unskip~\cite{1544700:26175285}.  The systematic building of Grover's Oracle for large-scale algorithms is a very important area of research\unskip~\cite{1544700:26157047}.  Therefore, there is a compelling reason to create a general algorithmic approach that a software development kit (SDK) could abstract down into easy-to-use interfaces for software developers to use in the wildcard search problem area. The suggested strategy ought to meet this condition as well.

Begin by formulating the Grover's Oracle as Equation~(\ref{dfg-4affd3545fe2}). 
\let\saveeqnno\theequation
\let\savefrac\frac
\def\dispfrac{\displaystyle\savefrac}
\begin{eqnarray}
\let\frac\dispfrac
\gdef\theequation{1}
\let\theHequation\theequation
\label{dfg-4affd3545fe2}
\begin{array}{@{}l}O_{}\vert x\rangle=\;\left\{\begin{array}{l}\;\;\;\vert x\rangle\;\;if\;x\;\neq\;w\\-\vert x\rangle\;\;if\;x\;=\;w\end{array}\right.\end{array}
\end{eqnarray}
\global\let\theequation\saveeqnno
\addtocounter{equation}{-1}\ignorespaces 
The Oracle operator would flip the sign of the target state. Putting the Oracle operator in matrix form in the computational basis yields Equation~(\ref{dfg-5b93e17b19b6})
\let\saveeqnno\theequation
\let\savefrac\frac
\def\dispfrac{\displaystyle\savefrac}
\begin{eqnarray}
\let\frac\dispfrac
\gdef\theequation{2}
\let\theHequation\theequation
\label{dfg-5b93e17b19b6}
\begin{array}{@{}l}O_{}=\;\begin{bmatrix}(-1)^{f(0)}&0&\cdots&0\\0&\ddots&0&\vdots\\\vdots&0&\ddots&0\\0&\dots&0&(-1)^{f(2^{i}-1)}\end{bmatrix}\end{array}
\end{eqnarray}
\global\let\theequation\saveeqnno
\addtocounter{equation}{-1}\ignorespaces 
where $i$ is the number of qubits and $f(x)$ equals $1$ if $x$ equals $w$ and $0$ if $x$ not equal to $w$.

The quantum circuit would begin by initializing the qubits into superposition states $|s\rangle$ using Hadamard gates $H$ as given in Equation~(\ref{dfg-d2a12dce789a})
\let\saveeqnno\theequation
\let\savefrac\frac
\def\dispfrac{\displaystyle\savefrac}
\begin{eqnarray}
\let\frac\dispfrac
\gdef\theequation{3}
\let\theHequation\theequation
\label{dfg-d2a12dce789a}
\begin{array}{@{}l}H^{\otimes i}\vert0\rangle^{i}\;=\;\frac1{\sqrt{2^{i}}}\sum_{j=0}^{2^{i}-1}\vert j\rangle=\vert s\rangle\end{array}
\end{eqnarray}
\global\let\theequation\saveeqnno
\addtocounter{equation}{-1}\ignorespaces 
The Grover's diffusion operator $D$, specified in Equation~(\ref{dfg-d571b8066631}), would be used to amplify the amplitude of the targeted state from the Oracle,
\let\saveeqnno\theequation
\let\savefrac\frac
\def\dispfrac{\displaystyle\savefrac}
\begin{eqnarray}
\let\frac\dispfrac
\gdef\theequation{4}
\let\theHequation\theequation
\label{dfg-d571b8066631}
\begin{array}{@{}l}D=2\vert s\rangle\!\langle s\vert - \mathbb{I}\end{array}
\end{eqnarray}
\global\let\theequation\saveeqnno
\addtocounter{equation}{-1}\ignorespaces 
where $\mathbb{I}$ is the identity matrix.

The entire quantum circuit would have $i$ qubit inputs that are initialized to zero and then individually coupled to Hadamard gates.  The Oracle-diffusion operator combination would be repeated roughly $(2^{i}/m)^{1/2}$ times for $m$ target states \unskip~\cite{1544700:26167079}.  The measurement would be taken at the end of the circuit.

\subsection*{Approach}The method is unique in that it converts wildcard search terms into the appropriate corresponding encoded expression that will be utilized to create the phase Oracle quantum circuit. To continue scaling the efficiency of quantum circuit construction, dependent SDK such as Qiskit\unskip~\cite{1544700:26166292}, Pennylane \unskip~\cite{1544700:26166568}, Braket\unskip~\cite{1544700:26166696} or Q\#\unskip~\cite{1544700:26166434} should abstract away the generation of the Grover's operators.  This paper's solution would utilize PKRM (Pseudo-Kronecker Reed Muller) synthesis\unskip~\cite{1544700:26167067} implemented in the tweedledum library\unskip~\cite{1544700:26167062}, which is an abstracted API (application programming interface) as part of Qiskit's PhaseOracle class\unskip~\cite{1544700:26167060} that accepts DIMAC CNF format\unskip~\cite{1544700:26167177} or logical expression.  The latter would be fed into the API.  This is to ensure that the experiments are focused on verifying the feasibility of using wildcard search in conjunction with Oracle as a database \unskip~\cite{1544700:26175289} and performing unstructured search using encoded search words.  PKRM synthesis is a procedure that would enable quantum circuit to be constructed programmatically based on the defined input. The logical statements containing the database and the wild card search phrases would be defined as the input for this work. Once more, the goal of the research is to enhance and streamline the creation of quantum circuits for the wildcard search domain using a scalable, programmable methodology.

The processing methodology for the three kinds of wildcard search would vary, with substring search being the most complicated. Prior to delving into the conversion process, it would be prudent to examine the data loading mechanism, which is a precondition for wildcard search.

\subsubsection*{Data set loading and encoding}Alphanumeric would make up the majority of the data in a software application used in practice.  Each character needs to be turned into a \textit{binary segment} before being put back together to form a \textit{binary entity}.  According to Equation~(\ref{dfg-dd7918f35544}), the encoding function (E) consumes the alphanumeric character and produces a \textit{binary segment} with a bit length greater than or equal to one.  The \textit{binary entity} would be produced by combining all of the \textit{binary segments} from one to n.
\let\saveeqnno\theequation
\let\savefrac\frac
\def\dispfrac{\displaystyle\savefrac}
\begin{eqnarray}
\let\frac\dispfrac
\gdef\theequation{5}
\let\theHequation\theequation
\label{dfg-dd7918f35544}
\begin{array}{@{}l}\left[E\left(a_1\right)\right],...,\lbrack E\left(a_{n-1}\right)\rbrack,\lbrack E\left(a_n\right)\rbrack\;for\;n\;\in\mathbb{Z},\;n\geq1\end{array}
\end{eqnarray}
\global\let\theequation\saveeqnno
\addtocounter{equation}{-1}\ignorespaces 
Consider a straightforward example where the character ``a'' is encoded in binary segment as ``0'' and the character ``b'' is encoded as ``1''. The binary representation of the character ``aba'' would be ``010'', with the first place ($[E(a_{1})]$ from Equation~(\ref{dfg-dd7918f35544})) representing ``a'' and the second position $([E(a_{2})])$ representing ``b''.  In the third place $([E(a_{3})])$, the ``0'' would be ``a''. 

A single bit in the binary segment is utilized in this instance to represent the letters ``a'' and ``b'', but more bits might be used to create longer, more complex alphanumeric characters and words. The condition would be to make sure that each binary segment within the binary entity is the same length.  

The encoded single data string is represented by one binary entity. To load all of the data strings into the quantum circuit, the binary entity for each string would be concatenated with an exclusive OR operator. The entire expression is known as a \textit{binary entity set}.

The process of creating the binary entity set would make it possible to load the encoded data into the phase Oracle and, ultimately, to perform amplitude amplification, which amplifies only the amplitudes of the qubit states that satisfy the binary entity set out of all possible superpositioned qubit states. The search phrases would then be concatenated with the expression for the imported data using the AND operator, depending on the wildcard search type. The goal is to identify the state vectors from all possible superpositioned states that first satisfy the logical criteria placed in the supplied data. The Oracle would behave like a database with loaded data contained in superpositioned state vectors. The loaded data's state vectors would then be filtered using the logical constraints of a wildcard search, and the desired states would be marked for amplitude amplification. The crucial factor of the approach would be the search term encoding.

\subsubsection*{Prefix search}Similar to how data is loaded, the prefix search word is also converted from alphanumeric to binary from left to right and character by character. The prefix search that would satisfy the objective would be ``ab*'' for instance, if the goal is to search for ``aba'' in addition to ``abb''. The binary entity conversion would produce ``01'' with ``0'' as the binary segment in location $[E(a_{1})]$ and ``1'' as the binary segment in place $[E(a_{2})]$. The ``*'' symbol will end any additional encoding.

\subsubsection*{Suffix search}The suffix search would use the opposite encoding order from the prefix search. Alphanumeric search terms would be entered into the phase oracle expression from right to left and would be encoded from right to left. For instance, the suffix search would be ``*bb'' if you wanted to look for ``bbb'' as well as ``abb''. The result of converting to a binary entity would be ``11'', with the ``1''s on the right and left representing $[E(a_{n})]$ and $[E(a_{n-1})]$, respectively.  The remaining binary segment would be empty in the binary entity.

\subsubsection*{Substring search}Prefix and suffix searches match the query phrase to each data string being searched; however, substring searches shift the query term by one character to look for the data string's substring. This shift and match operation will keep going until the last character of the query phrase matches the last character of the data string. For instance, while looking for ``ba'' in ``abbaa'', the data term substring ``ab'' will first be compared to the search word ``ba'', after which the data term will shift by one character to become ``bb'' and then be compared to ``ba'' once more. This method will keep going until either ``abbaa'' matches the search term ``ba'' or both strings' last characters are fulfilled.

As opposed to linear run time match and shift. To enable simultaneous searching, the search phrases could be transformed to binary representation. It would be necessary to transform the search word to a binary entity first. The search word is then concatenated after shifting by a bit segment. Up till the search word reaches the end of the binary entity, this shift and concatenate procedure will be repeated. In the previous example, the search word would be translated to the binary entity ``10'', while ``abbaa'' would be converted to the binary entity ``01100''. The suggested method is to build a logical expression for a substring search of the search phrase ``10'' based on the following expression.
\let\saveeqnno\theequation
\let\savefrac\frac
\def\dispfrac{\displaystyle\savefrac}
\begin{eqnarray}
\let\frac\dispfrac
\gdef\theequation{6}
\let\theHequation\theequation
\label{dfg-8057e59d5878}
\begin{array}{@{}l}\begin{array}{l}\lbrack E\left(a_1\right)\rbrack,\lbrack E\left(a_2\right)\rbrack\;\;OR\;\lbrack E\left(a_1\right)\rbrack,\lbrack E\left(a_2\right)\rbrack\;\;\;OR\;\\...\;OR\;\lbrack E\left(a_{n-2}\right)\rbrack,\lbrack E\left(a_{n-1}\right)\rbrack\;OR\;\lbrack E\left(a_{n-1}\right)\rbrack,\lbrack E\left(a_n\right)\rbrack\;\;\;\;\end{array}\end{array}
\end{eqnarray}
\global\let\theequation\saveeqnno
\addtocounter{equation}{-1}\ignorespaces 
With $n$ - (binary segment's bit length) + 1 terms.

Using the logical OR operator, this binary entity set would be combined with the binary entity sets from other search queries.

\subsubsection*{Phase Oracle Algorithm}The following figure shows the algorithm for creating the phase Oracle expression from encoding and merging binary entity sets.  The algorithm would produce the logical statement needed by the PKRM synthesis, which would transform the statement into the functional quantum circuit.

\begin{lstlisting}[style=listing_style,caption={Algorithm for transforming the logical statements needed to generate the Phased Oracle quantum circuit from the classical database and the wildcard search phrases}]
Iterable data set [@\textit{D}@]
DataExpression = []
[@\textbf{for }@][@\textit{d}@] in [@\textit{D}@][@\textbf{ do}@] 
  DataExpression.append(Encode([@\textit{d}@]))
string OracleExpression = DataExpression.join('[@{\textasciicircum}@]')
Iterable search terms [@\textit{S}@]  
SearchExpression = []
[@\textbf{for }@][@\textit{s}@] in [@\textit{S}@][@\textbf{ do}@]
  Switch([@\textit{s}@])
    [@\textbf{case}@] PrefixSearch [@\textbf{do}@]
      SearchExpression.append(PrefixEncode([@\textit{s}@]))
    [@\textbf{case}@] SuffixSearch [@\textbf{do}@]
      SearchExpression.append(SuffixEncode([@\textit{s}@]))
    [@\textbf{case}@] SubstringSearch [@\textbf{do}@]
      SearchExpression.append(SubstringEncode([@\textit{s}@]))
 
string OracleSearchExpression = SearchExpression.join('|')
[@\textbf{return}@] OracleExpression + '[@\&@]' + OracleSearchExpression

\end{lstlisting}
The encoding functions make reference to the aforementioned kinds of wildcard searches and the accompanying entity set building processes.

\subsubsection*{Quantum circuit}As described in the Background section, the phase Oracle would be inserted into Grover's operator. Prior to connecting to the Grover's operator, the initialized zero states would be linked to their corresponding Hadamard gates. Due to the nature of Grover's quantum circuit, the output state being measured would contain qubits in the reverse manner.

\subsection*{Literature review}Numerous publications\unskip~\cite{1544700:26167179}\unskip~\cite{1544700:26175289} explore the mechanism of loading classical data into the quantum circuit via phase input or Grover's Oracle, as well as the viability of multi-object search\unskip~\cite{1544700:26167079}.  The research's novelty resides in the technical viability of programmatically converting unstructured data and wildcard search phrases that are employed by classical computers into logical statements that are then ingested by PKRM synthesis and systemically generate a functional quantum circuit. This would make creating quantum circuits easier and increase their likelihood of being adopted by the engineering community. The research is founded on the papers cited as references.
    
\section*{Methods}
Three kinds of matching outcomes would be tested: one matching result, two matching results, and no matching result. Wildcard search would be used for all searches. The experiments with no matching results are meant to serve as the experiment's control. Additionally, the experiments would be performed on two different kinds of quantum computers: trapped ion and super conducting types.  

The needed data must be specified and sifted through all potential states before the quantum circuit can be created. The wildcard search phrase should then be defined based on the loaded states after that. In Table~\ref{tw-a865721b47fa}, the ``bit encoding'' column lists every potential state for a processor with three qubits. Four states were selected from the ``loaded data'' to be utilized in the wildcard search. The state for one match or two matches is displayed in the ``wildcard search'' column.

\begin{table*}[!htbp]
\caption{{Qubit encoding truth table} }
\label{tw-a865721b47fa}
\def\arraystretch{1}
\ignorespaces 
\centering 
\begin{tabulary}{\linewidth}{LLL}
\hline \cAlignHack bit encoding & \cAlignHack loaded data & \cAlignHack wildcard search\\
\hline 
\cAlignHack 000 &
  \cAlignHack Y &
  \cAlignHack one\_match\\
\cAlignHack 001 &
  \cAlignHack  &
  \cAlignHack \\
\cAlignHack 010 &
  \cAlignHack Y &
  \cAlignHack two\_match\\
\cAlignHack 011 &
  \cAlignHack Y &
  \cAlignHack two\_match\\
\cAlignHack 100 &
  \cAlignHack  &
  \cAlignHack \\
\cAlignHack 101 &
  \cAlignHack  &
  \cAlignHack \\
\cAlignHack 110 &
  \cAlignHack  &
  \cAlignHack \\
\cAlignHack 111 &
  \cAlignHack Y &
  \cAlignHack \\
\hline 
\end{tabulary}\par 
\end{table*}
The next phase would be creating the quantum circuits, specifically the Oracle, with the appropriate search target already in place.

\subsection*{Generating quantum circuits}The Phase Oracle would be constructed first, as it would load data before performing a wildcard search. Consequently, there would be two components to the Oracle: the data set loading and the actual wildcard search components.

\subsubsection*{Phase Oracle}According to Table~\ref{tw-a865721b47fa}, the data state would be encoded using logical operators. The NOT operator would be used in front of the alphanumeric letter denoting the position of the qubit in the binary entity to signify the 0 state qubit. An exclusive OR operator is used to separate binary entities before combining them to produce the binary entity set.  

The wildcard search word would be the second part of the Oracle. The wildcard search phrase for the one-match experiments is ``*1*''. This would find any data with the second qubit set to ``1''. The wildcard search phrase for the two-matches experiments is ``01*''. This would find any data containing ``0'' in the first qubit and ``1'' in the second. Table~\ref{tw-d8796af84c15} shows the mapping of search phrase to experiment intended outcome.

\begin{table*}[!htbp]
\caption{{Wild card search term mapping to expected outcome} }
\label{tw-d8796af84c15}
\def\arraystretch{1}
\ignorespaces 
\centering 
\begin{tabulary}{\linewidth}{LL}
\hline \cAlignHack Expected outcome & \cAlignHack Wild card search term\\
\hline 
\cAlignHack no\_match (control) &
  \cAlignHack 1 0 *\\
\cAlignHack one\_match &
  \cAlignHack * 1 *\\
\cAlignHack two\_match &
  \cAlignHack 0 1 *\\
\hline 
\end{tabulary}\par 
\end{table*}
Finally, using an ``AND'' operator, the loaded data and search terms would be combined. The Qiskit SDK would be used to help generate the circuit gates shown in Figures~\ref{f-7a09a41e6ecc} and~\ref{f-0b28532d7882}.

\subsubsection*{Full circuit}For amplitude amplification of the desired states, the generated Oracle would be connected to Grover's operator circuit. This would adhere to the speculative algorithm mentioned in the Background section. Then, the states of the three input qubits would be set to ``0'', and each qubit would be connected to a matching Hadamard Gate. The circuit would then establish a connection with the Grover's operator from above.  For a one-match circuit and a two-match circuit, respectively, see Figure~\ref{f-7a09a41e6ecc} and Figure~\ref{f-0b28532d7882}.

\bgroup
\fixFloatSize{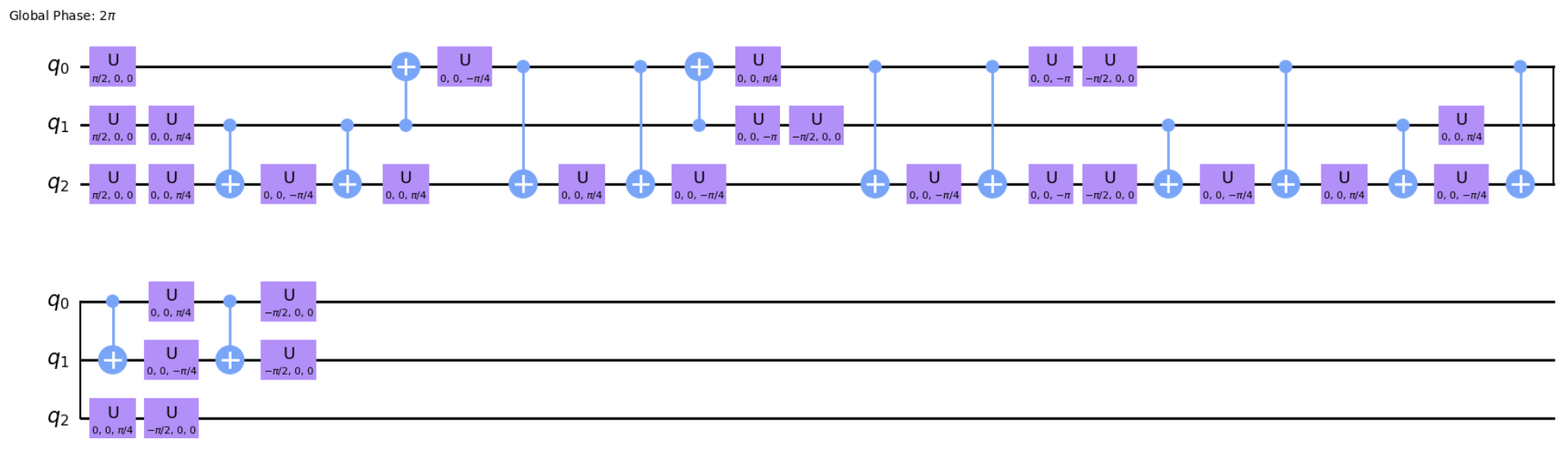}
\begin{figure*}[!htbp]
\centering \makeatletter\IfFileExists{images/transpiled_full_circuit.png}{\includegraphics{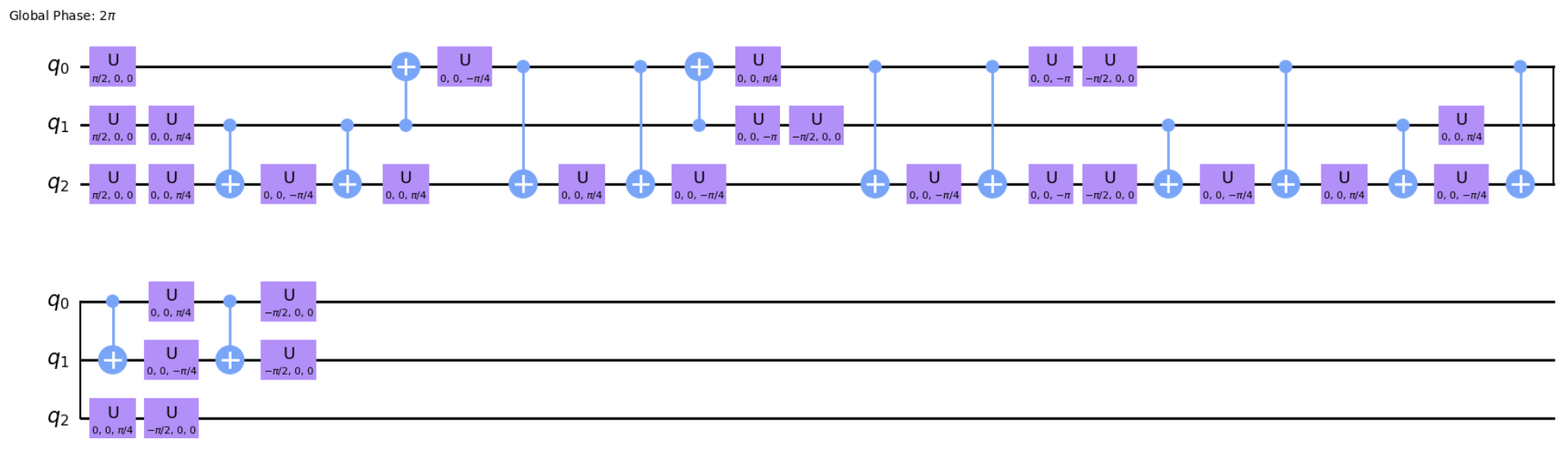}}{}
\makeatother 
\caption{{Quantum circuit for one match in 3 qubits}}
\label{f-7a09a41e6ecc}
\end{figure*}
\egroup

\bgroup
\fixFloatSize{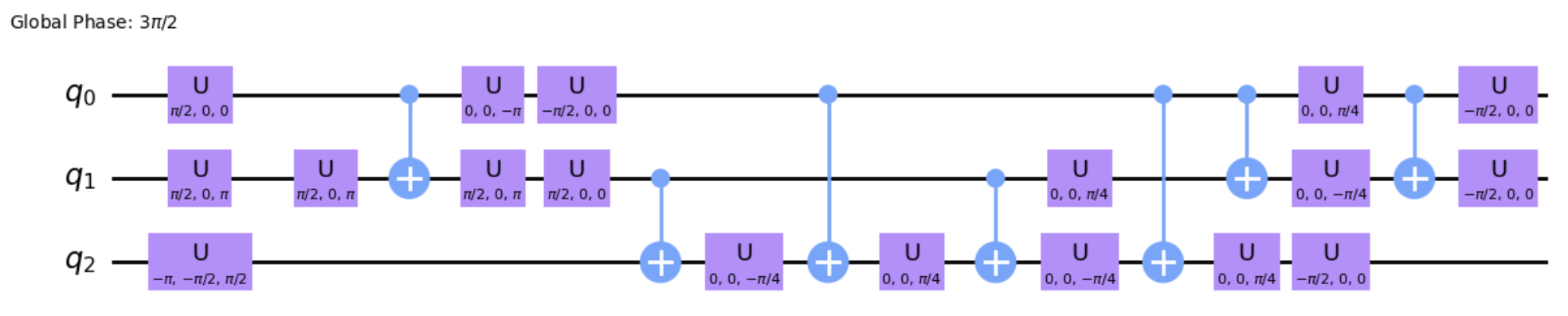}
\begin{figure*}[!htbp]
\centering \makeatletter\IfFileExists{images/transpile_full_circuit_decompose.png}{\includegraphics{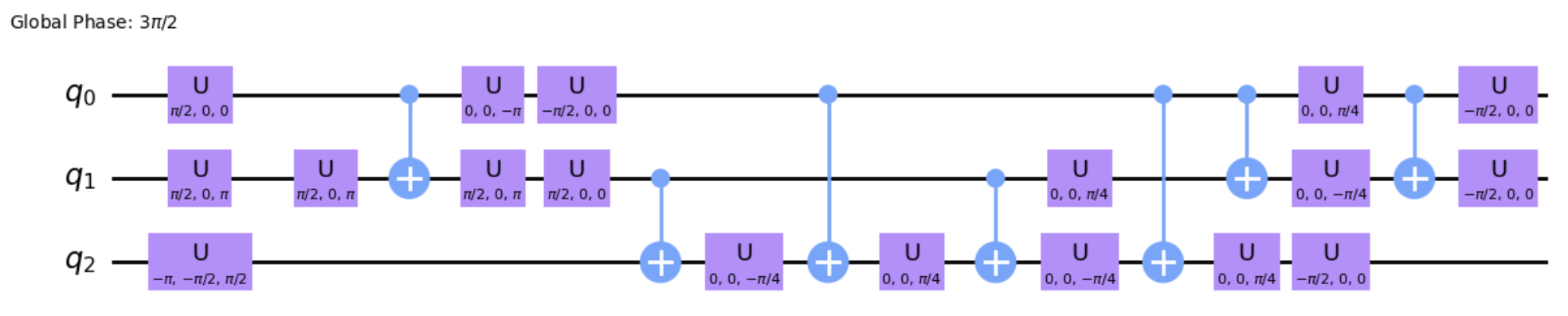}}{}
\makeatother 
\caption{{Quantum circuit for two matches in 3 qubits}}
\label{f-0b28532d7882}
\end{figure*}
\egroup

\subsection*{State vector}Bloch Spheres could be employed to represent the state vectors for the circuits' simulated outputs prior to measurement. Figure~\ref{f-b658738214bb} shows the Bloch Sphere for one-match quantum circuits, whereas Figure~\ref{f-d2e2668120ee} shows it for two-match quantum circuits. These visual displays of the simulated state vector before to measurement would be compared to the experimental findings and be further examined in the Discussion section.

\bgroup
\fixFloatSize{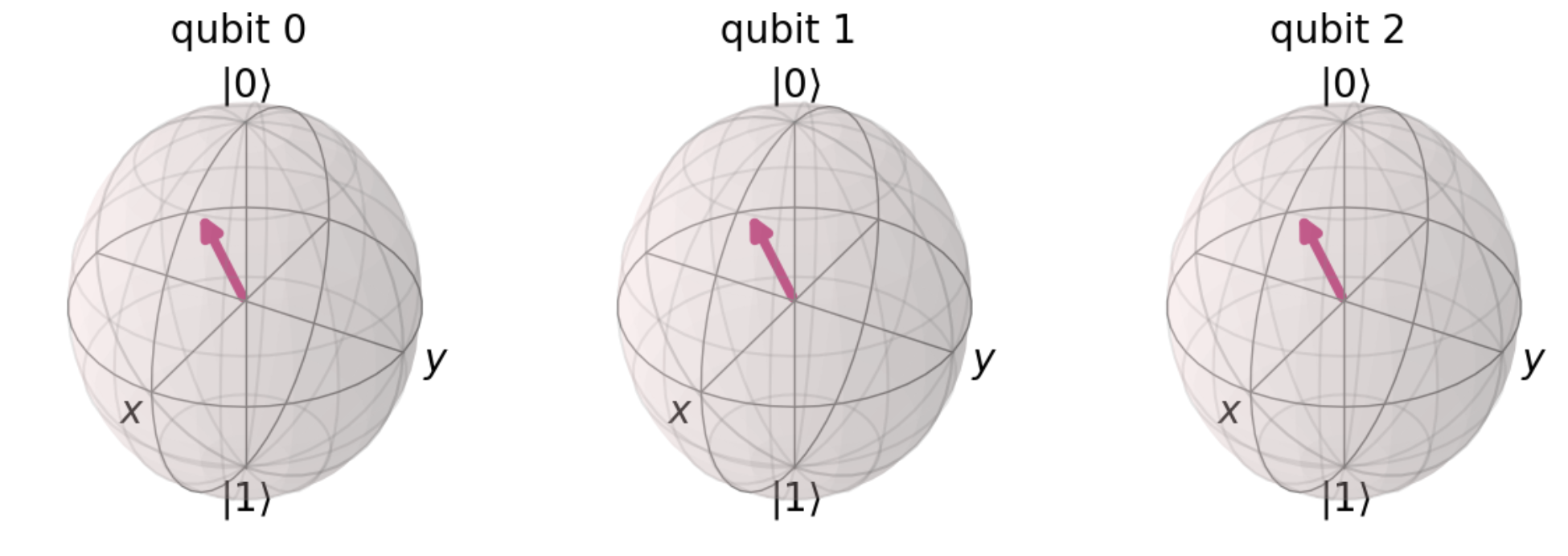}
\begin{figure}[!htbp]
\centering \makeatletter\IfFileExists{images/screen-shot-2022-10-05-at-11-u03-u58-pm.png}{\includegraphics{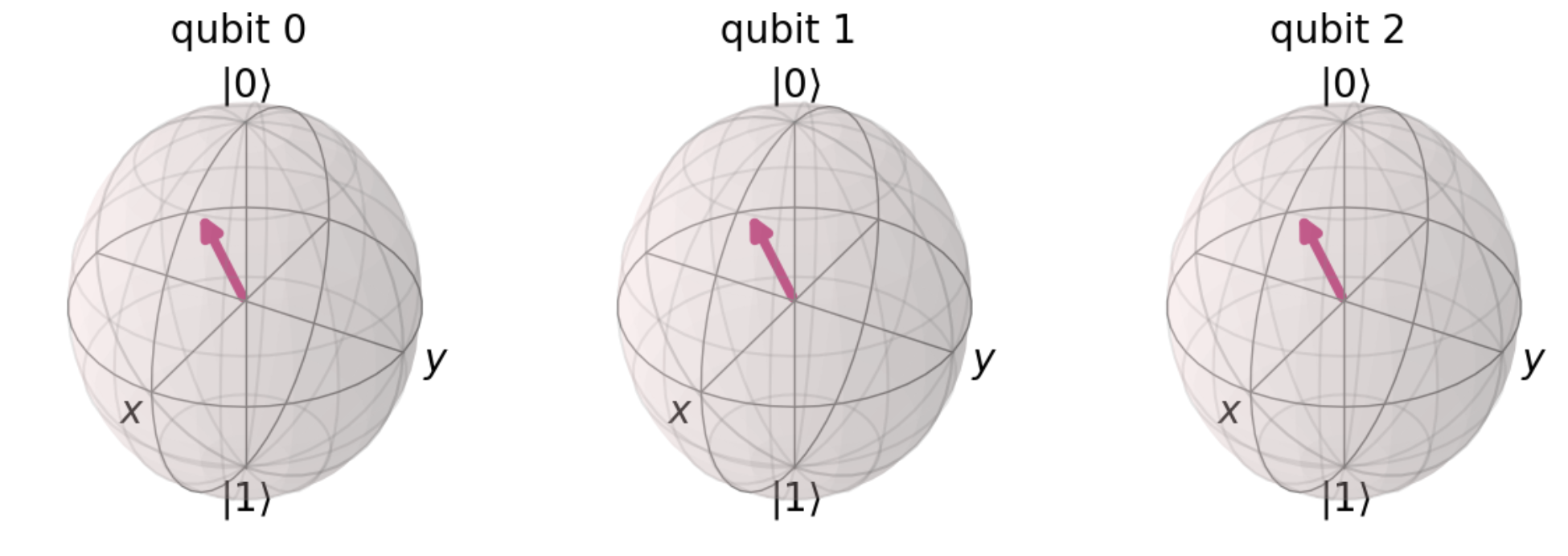}}{}
\makeatother 
\caption{{Bloch Sphere for one match on 3 qubits }}
\label{f-b658738214bb}
\end{figure}
\egroup

\bgroup
\fixFloatSize{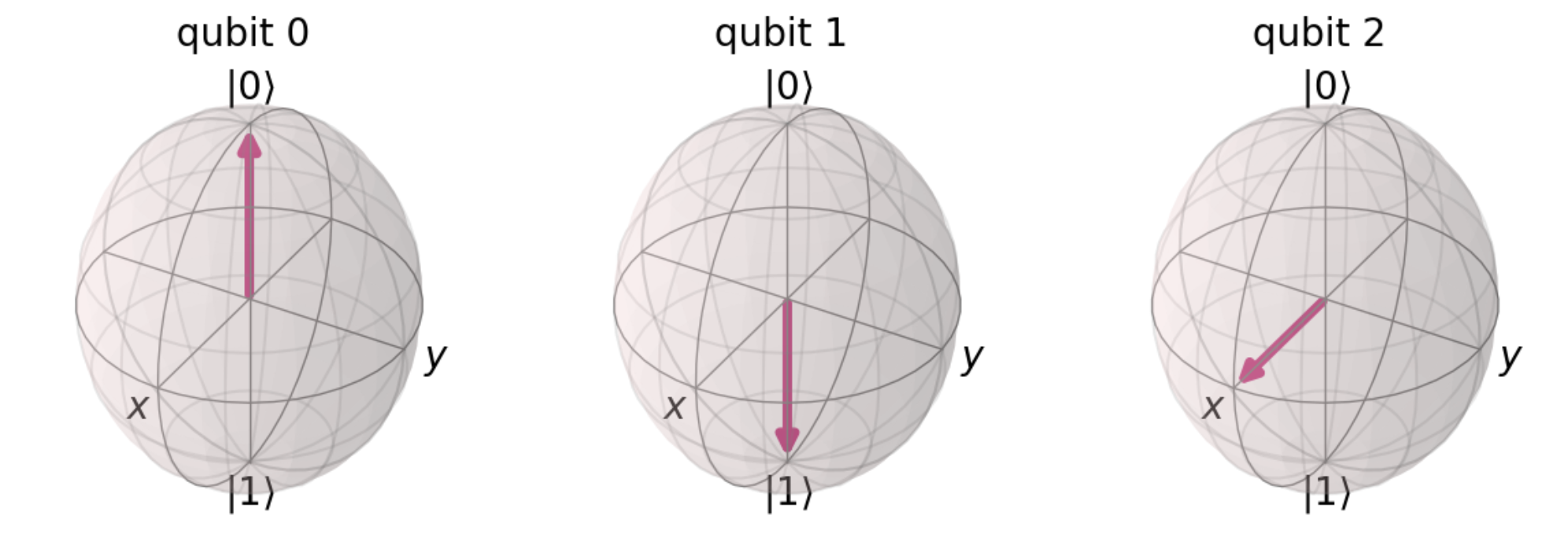}
\begin{figure}[!htbp]
\centering \makeatletter\IfFileExists{images/bloch_sphere.png}{\includegraphics{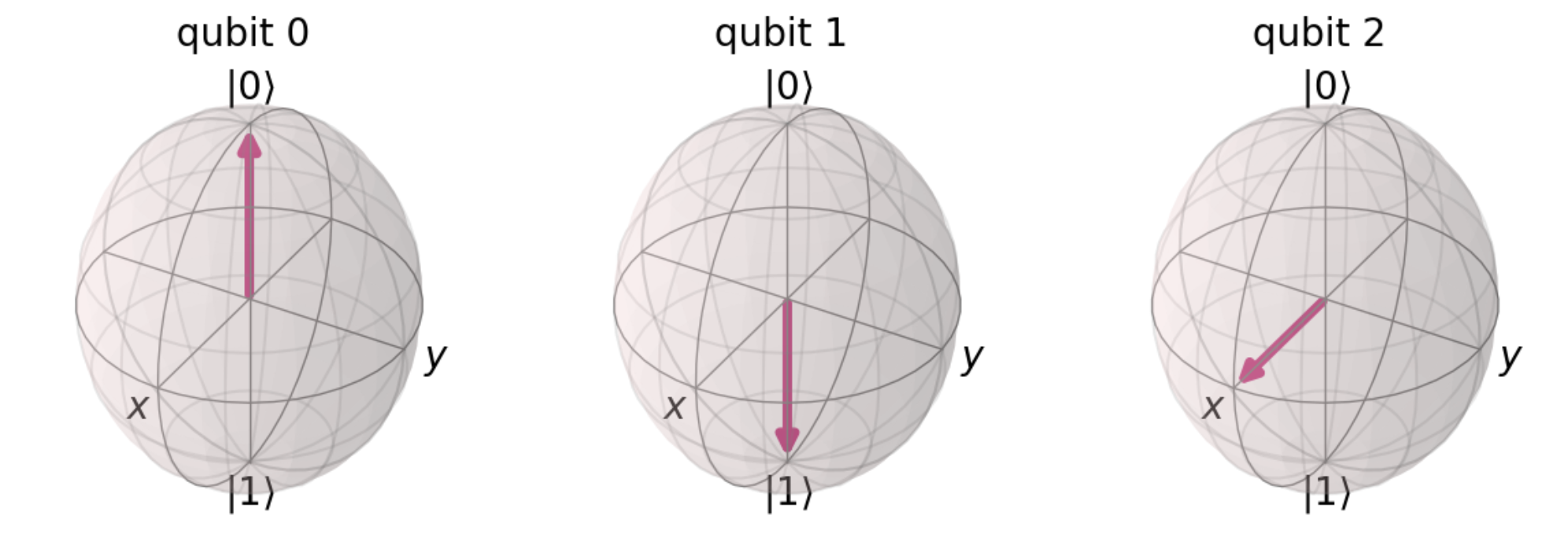}}{}
\makeatother 
\caption{{Bloch Sphere for two matches on 3 qubits}}
\label{f-d2e2668120ee}
\end{figure}
\egroup

\subsection*{Running on NISQ processors}On a superconducting quantum computer and a trapped ion-based quantum computer, respectively, the two circuits of Figure~\ref{f-7a09a41e6ecc} and Figure~\ref{f-0b28532d7882} would be executed. The processors' outputs would be measured and reported in accordance. The Appendix contains a description of the processor specifications. On each processor, the experiment would be run six times for the no-match (control), one-match, and two-matches scenarios. 
    
\section*{Results}
Figure~\ref{f-0568b26d39db} depicts the metrics of the comparison between trapped ion type and superconducting qubit quantum processors for a two-matches scenario on a single run. The probabilities were standardized in this manner. Because of the nature of Grover's diffusion operator, as discussed in the Background section, the state qubits are inverted.

\bgroup
\fixFloatSize{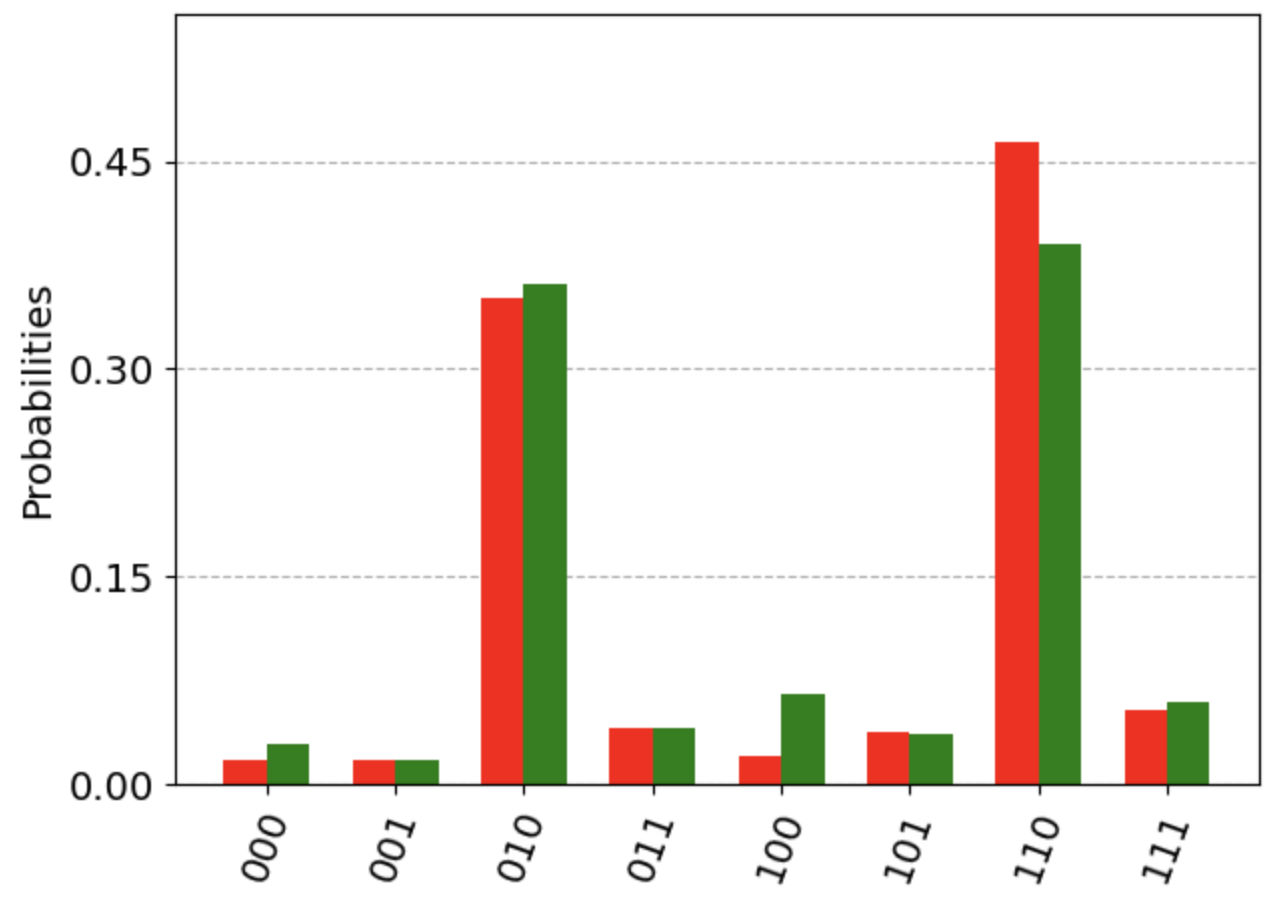}
\begin{figure*}[!htbp]
\centering \makeatletter\IfFileExists{images/2match_v2.png}{\includegraphics{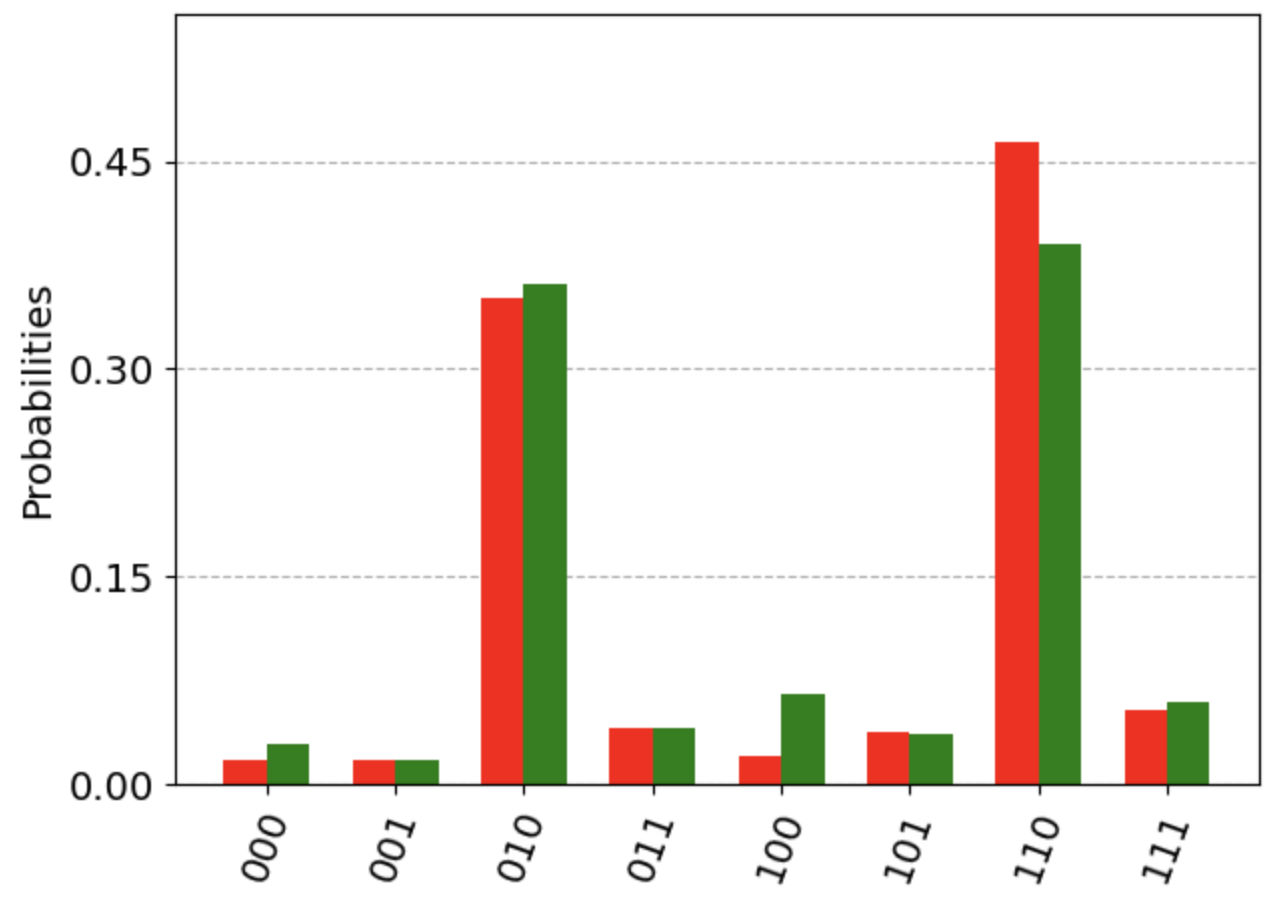}}{}
\makeatother 
\caption{{Results from two 3 qubit processors in a sample two-matches scenario (red being trapped ion type and green being superconducting qubits). The states are situated on the X-axis with the qubit order reversed. }}
\label{f-0568b26d39db}
\end{figure*}
\egroup
Figure~\ref{f-35d2a025fb28} shows the metrics for comparing the performance of two different quantum computing systems for the outcome of a trial in the one-match scenario. The green (right) bars show the output from superconducting qubits, whereas the red (left) bars show the trial output from the trapped ion quantum computer.

\bgroup
\fixFloatSize{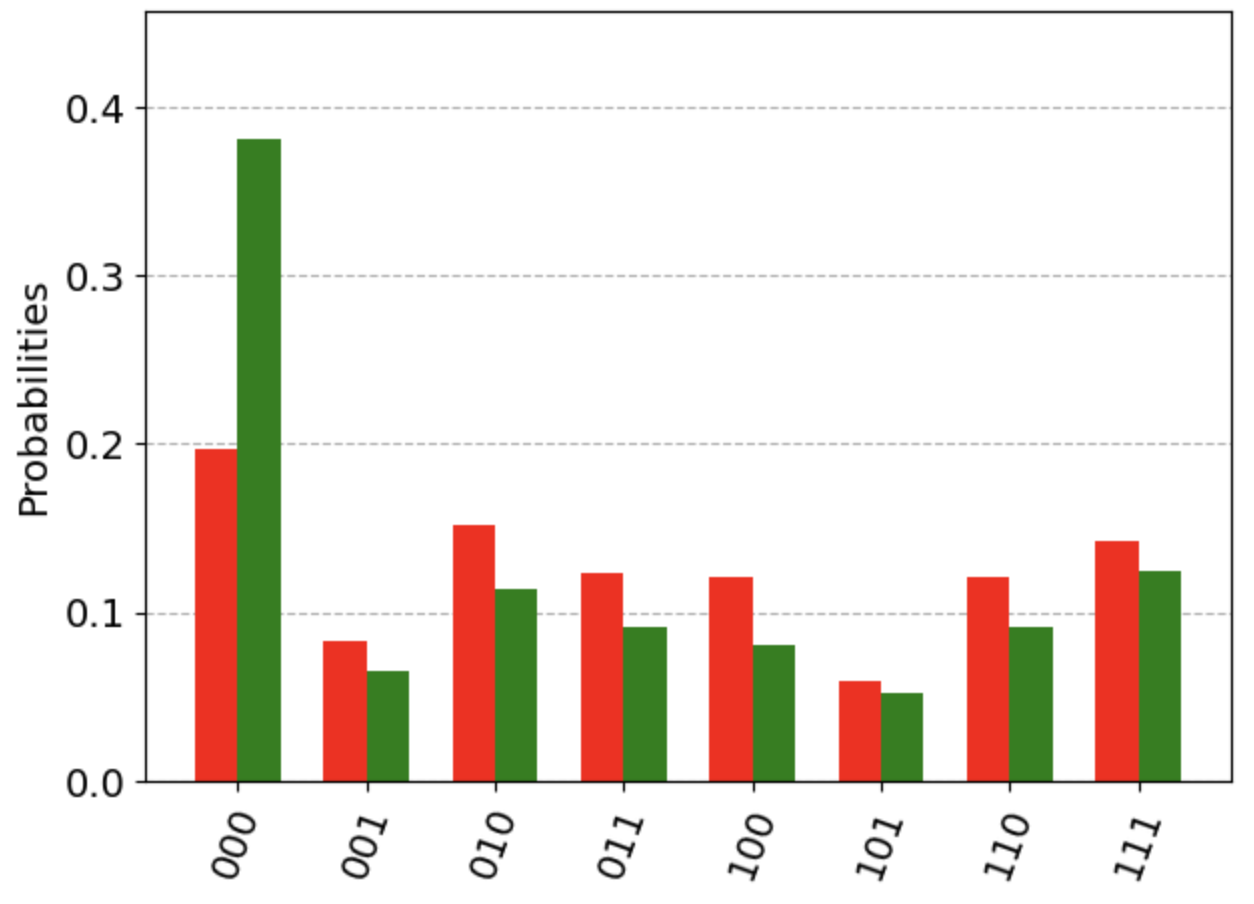}
\begin{figure*}[!htbp]
\centering \makeatletter\IfFileExists{images/1match_v2.png}{\includegraphics{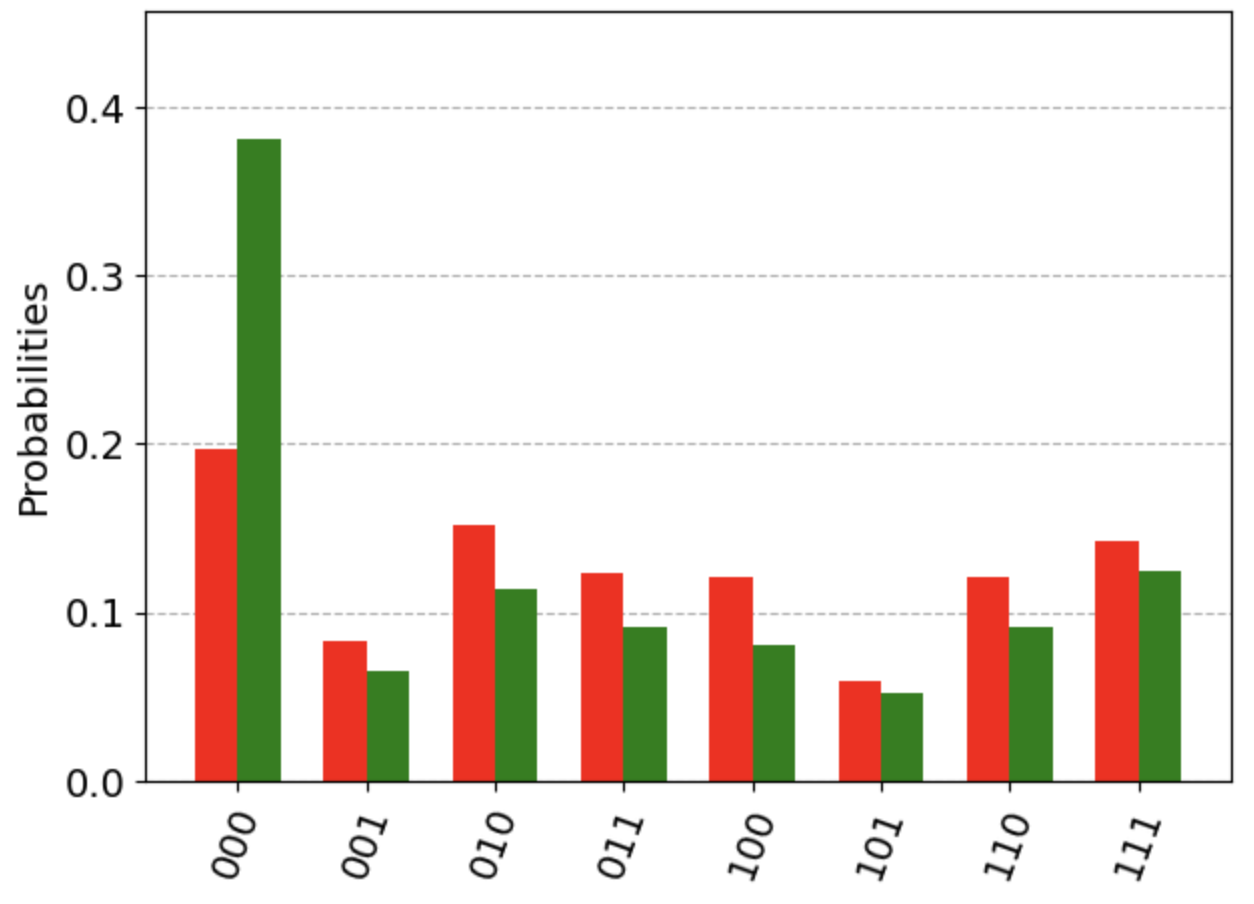}}{}
\makeatother 
\caption{{Results from two 3 qubit processors in a sample one-match scenario (red being trapped ion type and green being superconducting qubits). The states are situated on the X-axis with the qubit order reversed.}}
\label{f-35d2a025fb28}
\end{figure*}
\egroup
Table~\ref{tw-741cc11c763e} shows the results of all six trials run on a trapped ion type processor. For the one-match and no-match scenarios, the state with the highest probability is recorded, and the states with the top two highest probabilities are recorded for the two-match scenario. The table will show the consistency of the output and will be thoroughly examined in the Discussion section.

\begin{table*}[!htbp]
\caption{{Tabulated experimental findings for processor adopting trapped ions, with maximum values recorded. The values between brackets represent the probabilities of the state.} }
\label{tw-741cc11c763e}
\def\arraystretch{1}
\ignorespaces 
\centering 
\begin{tabulary}{\linewidth}{p{\dimexpr.11220000000000002\linewidth-2\tabcolsep}p{\dimexpr.2881\linewidth-2\tabcolsep}p{\dimexpr.20600000000000005\linewidth-2\tabcolsep}p{\dimexpr.3937\linewidth-2\tabcolsep}}
\hline Trials & 0 match (control) & 1 match & 2 matches\\
\hline 
\cAlignHack 1 &
  \cAlignHack 101 (0.148) &
  \cAlignHack 000 (0.179) &
  \cAlignHack 011 (0.464), 010 (0.351)\\
\cAlignHack 2 &
  \cAlignHack 001 (0.165) &
  \cAlignHack 000 (0.188) &
  \cAlignHack 011 (0.468), 010 (0.365)\\
\cAlignHack 3 &
  \cAlignHack 111 (0.161) &
  \cAlignHack 000 (0.197) &
  \cAlignHack 011 (0.466), 010 (0.367)\\
\cAlignHack 4 &
  \cAlignHack 101 (0.164) &
  \cAlignHack 000 (0.182) &
  \cAlignHack 011 (0.454), 010 (0.381)\\
\cAlignHack 5 &
  \cAlignHack 111 (0.190) &
  \cAlignHack 000 (0.175) &
  \cAlignHack 011 (0.456), 010 (0.370)\\
\cAlignHack 6 &
  \cAlignHack 111 (0.187) &
  \cAlignHack 000 (0.215) &
  \cAlignHack 011 (0.431), 010 (0.387)\\
\hline 
\end{tabulary}\par 
\end{table*}
Table~\ref{tw-9330e477d705} contains the experimental results from the identical six tests performed on a superconducting qubit type quantum processor.

\begin{table*}[!htbp]
\caption{{Tabulated experimental findings for processor adopting superconducting qubits, with maximum values recorded. The values between brackets represent the probabilities of the state.} }
\label{tw-9330e477d705}
\def\arraystretch{1}
\ignorespaces 
\centering 
\begin{tabulary}{\linewidth}{LLLL}
\hline Trials & 0 match (control) & 1 match & 2 match\\
\hline 
\cAlignHack 1 &
  \cAlignHack 110 (0.165) &
  \cAlignHack 000 (0.210) &
  \cAlignHack 011 (0.267), 010 (0.208)\\
\cAlignHack 2 &
  \cAlignHack 111 (0.194) &
  \cAlignHack 000 (0.284) &
  \cAlignHack 011 (0.263), 010(0.255)\\
\cAlignHack 3 &
  \cAlignHack 110 (0.176) &
  \cAlignHack 000 (0.381) &
  \cAlignHack 011 (0.337), 010 (0.255)\\
\cAlignHack 4 &
  \cAlignHack 010 (0.199) &
  \cAlignHack 000 (0.380) &
  \cAlignHack 011 (0.391), 010 (0.361)\\
\cAlignHack 5 &
  \cAlignHack 110 (0.187) &
  \cAlignHack 000 (0.186) &
  \cAlignHack 011 (0.418), 010 (0.323)\\
\cAlignHack 6 &
  \cAlignHack 110 (0.188) &
  \cAlignHack 000 (0.321) &
  \cAlignHack 011 (0.292), 010 (0.406)\\
\hline 
\end{tabulary}\par 
\end{table*}

\section*{Discussion}
When the output from both types of quantum computers is compared to the theoretical vector state from Figure~\ref{f-d2e2668120ee} in the two-match situation shown in Figure~\ref{f-0568b26d39db}, the experimental findings are similar to what is predicted theoretically. Particularly, the two bar graphs clearly show high probability for the ``010'' and ``011'' qubit entities for the two-match situation shown in Table~\ref{tw-a865721b47fa}. The two-match column of the tabulated results from all six experiments carried out on the trapped ion quantum computer, as shown in Table~\ref{tw-741cc11c763e}, made it abundantly evident that the two qubit entities have the top two highest probabilities among all possible states. Additionally, Table~\ref{tw-9330e477d705}'s two-match column from all six tests conducted on a superconducting qubit quantum computer lends weight to the conclusion. It should be noted that the outputs were intended to be condensed such that, for the two-match scenario, the top two states with the highest probabilities were displayed in the two tables, and that, for the one-match and zero-match scenarios, only the state with the highest probabilities was displayed. The goal is to make identifying consistency in the collected data easier. For instance, if the states with the highest probability were to consistently shift, as in the ``no-match'' scenario column in Tables~\ref{tw-741cc11c763e} and~\ref{tw-9330e477d705}, that would be a sign that no states matched the search criteria and the results were random, and vice versa.

Based on the truth table from Table~\ref{tw-a865721b47fa}, the one-match scenario looks for the qubit entity of ``000''. According to Figure~\ref{f-35d2a025fb28}, a superconducting qubit quantum processor typically has a greater capacity to determine the expected state than a trapped ion quantum processor. In addition, as demonstrated in Table~\ref{tw-9330e477d705}'s one-match column for all six trials, superconducting qubit processors tend to have greater state probabilities than trapped ion processors do for all six studies. The results also match those of the anticipated state vector displayed in Figure~\ref{f-b658738214bb}'s Bloch sphere.  Realizing that finding a scalable approach to systemically assess the matching findings is one of the research work's approval criteria is crucial. It would be possible for the software to identify the matching outcomes in NISQ processors effectively by running the experiments over a number of trials and selecting the states with the highest $K$-th probability that are consistent across all of the trials.  

In summary, the results of the experiments support the proposed strategy for three-qubit wildcard search. However, there are limits that must be addressed in future study projects.

\subsection*{Limitations}Quantum computers are very susceptible to noise, which influences the gap between actual and expected outcomes\unskip~\cite{1544700:26194379}. This is true for both types of processors. The quantum circuit's depth is correlated to the level of noise, which impacts the correctness of the outcome.

Due to an increase in the quantum circuit depth, the accuracy of the result would decline for bigger qubit states than three qubits. For NISQ processors, this is the current hardware limitation. There are methods \unskip~\cite{1544700:26194380}\unskip~\cite{1544700:26194403}  to reduce noise, though, and these might be investigated in the future. The wildcard substring search's viability would be further confirmed, and having larger qubit state would bring it one step closer to practical use.

\subsection*{Future work}Perform the experiments using larger qubit states ({\textgreater}10 qubits) and noise reduction techniques. Larger data sets and more complicated wild card search phrases would be possible because of the large qubit state. This would bring the approach suggested in this research one step closer to actual use by evaluating its viability in larger data sets.
    
\section*{Conclusions}
During the experimental period, there were no reports of any wildcard search issues being solved programmatically with 3 qubits of Grover's algorithm and a created Oracle comprising the input data and search terms. The difficulties of data loading and search term loading into the quantum circuit are not straightforward; therefore, the ability to perform wildcard search in addition to loading the data and the circuit in a linear amount of time could indicate quantum advantage for the problem space. Quantum computer noises would be one of the most significant obstacles to executing the proposed technology in practice. For the suggested strategy, the NISQ processors would not be able to manage more than three qubits. Future research would utilize additional qubits with deeper quantum circuits and continue to push the NISQ computers to their limits.
    
\section{List of abbreviations}
.

NISQ - Noisy intermediate-scale quantum

PKRM - Pseudo-Kronecker Reed Muller

AWS - Amazon web services

API - application programming interface

\section*{Declarations}

\subsection*{Availability of data and materials}The data sets used and/or analyzed during the current study are available from the corresponding author on reasonable request.

\subsection*{Funding}AWS Kumo 
    
\section*{Authors' contributions}
The theoretical technique was proposed by the author, who then tested it using NISQ processors as a service from AWS. Author created this paper after compiling the findings.
\section*{Acknowledgments}Not applicable

\bibliography{article}

\end{document}